\documentclass[onecolumn,showpacs,prd,groupedaddress,nofootinbib]{revtex4}

\usepackage{amsfonts,amsmath,amssymb,mathrsfs}
\usepackage{hyperref}
\usepackage{subfigure}
\usepackage{color}
\usepackage{graphicx}  
\usepackage{dcolumn}   
\usepackage{bm}        
\usepackage[english]{babel}

\def\a{\alpha}
\def\b{\beta}
\def\g{\gamma} \def\G{\Gamma}
\def\d{\delta} 

\def\z{\zeta}

\def\m{\mu}
\def\n{\nu}
\def\r{\rho}
\def\s{\sigma} \def\S{\Sigma }

\def\o{\omega} \def\O{\Omega}

\def\tx{\tilde{x}}
\def\tV{\tilde{V}}

\def\la{\langle}
\def\ra{\rangle}

\def\bea{\begin{equation}}
\def\eea{\end{equation}}
\def\beq{\begin{eqnarray}}
\def\eeq{\end{eqnarray}}
\def\ben{\begin{enumerate}}
\def\een{\end{enumerate}}

\def\p{\partial}
\def\ph{\phantom}
\def\oythree{O\left(y^3\right)} 
\def\oyfour{O\left(y^4\right)}
\def\oxone{O\left(x\right)}
\def\oxtwo{O\left(x^2\right)}
\def\oxthree{O\left(x^3\right)}
\def\oxfour{O\left(x^4\right)}
\def\oxfive{O\left(x^5\right)}
\def\oVtwo{O\left(V^2\right)}

\begin{document}

\title{Locally inertial null normal coordinates}
\author{Raf Guedens}
\affiliation{rafguedens@gmail.com}
\begin{abstract}
Locally inertial coordinates are constructed by carrying Riemann normal coordinates on a codimension two spacelike surface along the geodesics normal to it. Since the normal tangents are labelled by components with respect to a null basis, these coordinates are referred to as null normal coordinates. They are convenient in the study of local causal horizons. As an application, the coordinate system is used to specify a vector field that satisfies the Killing equation approximately in a small region and the Killing identity exactly on a single null geodesic. We also construct a vector field on a surface, starting from a vector at a given point on the surface. This construction may be regarded as a generalisation of the notion of Fermi-Walker transport.   
\end{abstract} 
\maketitle

\section{Introduction} 

Different problems call for different coordinate systems. For example, a fluid whose flow lines are hypersurface-orthogonal geodesics is suitably described in terms of Gaussian coordinates. These are defined as follows: On a (portion of a) spatial hypersurface $\O$ some arbitrary coordinates are placed. A point $r$ in the neighborhood of $\O$ lies on a unique timelike geodesic intersecting $\O$ orthogonally, at $q$ say. Then the Gaussian coordinates of $r$ consist of the  proper time from $q$ to $r$ along the geodesic, together with the spatial coordinates of $q$ on $\O$. 

Null normal coordinates (NNCs) are defined in a similar way, but starting from a codimension two spatial surface $\S$. At each point on $\S$, the tangent space orthogonal to $\S$ may be spanned by a pair of null vectors.
A point $r$ in the neighborhood of $\S$ again lies on a unique geodesic intersecting $\S$ orthogonally, at $q$ say. At this point $q$, the tangent to the geodesic will be a linear combination of the pair of null vectors. If the geodesic is parametrized such that $r$ lies at unit affine parameter, then the NNCs of $r$ are given by the coefficients of this linear combination, together with arbitrary coordinates of $q$ on $\S$. Accordingly, every null geodesic orthogonal to $\S$ is a coordinate curve. 

NNCs were introduced in \cite{Kay:1988mu, Iyer:1994ys} in the context of black hole horizons, where $\S$ was chosen to be a cross section of the horizon. However, from the definition above it is clear that NNCs may be constructed off any codimension two spatial surface in an arbitrary spacetime, as long as the surface can be covered with a single coordinate chart and the orthogonal geodesics do not intersect. Note also that the orthogonal null geodesics generate a pair of null hypersurfaces, each cut in half by the surface. This calls to mind the ``lightsheets" of the covariant entropy conjecture put forward in \cite{Bousso:1999xy}. 

Furthermore, the null surfaces to the past of $\S$ form part of the boundary of its past. Therefore, if $\S$ is very small, the null surfaces are part of a local causal horizon. Such local horizons have been employed to address the question whether gravitational field equations are a manifestation of the Clausius relation of thermodynamics \cite{Jacobson:1995ab, Guedens:2011dy}. 

In section \ref{LINNCs} we will construct NNCs starting from a particular kind of small surface $\S$, namely one that is generated by geodesics. More precisely, $\S$ is the image under the exponential map of a small part of a spatial, codimension two tangent space of a spacetime point $p$. Starting from such a surface has the added advantage that the NNCs can be made to be locally inertial, i.e. at $p$ the coordinate vectors are orthonormal\footnote{See the remark below Eqn. (\ref{NNC}) for a qualification of this statement.} and the Christoffel symbols vanish. It will also be shown that the coordinates we introduce are, in some sense, the simplest possible to have the properties of being null normal and locally inertial.               

The construction of the coordinate system requires a pair of null vector fields orthogonal to the surface $\Sigma$. While in section \ref{LINNCs} these are specified with the aid of a Riemann normal coordinate system, such vector fields may equivalently be defined in a coordinate independent manner, via a generalisation of the notion of Fermi-Walker transport. We introduce this notion in the appendix. Starting from a given pair of null vectors at a point and transporting according the prescription given in the appendix uniquely retrieves the vector fields constructed in section \ref{LINNCs}. 

An important ingredient in the conjectured thermodynamic nature of gravity is a collection of local, approximate Killing vector fields. The components of these vector fields in locally inertial coordinates are such that boosts would be generated if the coordinates were globally inertial. Then the Killing equation is satisfied in the neighbourhood of a spacetime point, up to possible corrections linear in the locally inertial coordinates. 

However, this does not specify the vector fields uniquely. Using the coordinate system of section \ref{LINNCs}, we will show in section \ref{LKV} that it is possible to further specify the fields such that they satisfy the Killing identity\footnote{For a true Killing vector, the Killing identity $\nabla_a \nabla_b \xi_c = R^d{}_{abc} \xi_d$ obtains by taking covariant derivatives of the Killing equation $\nabla_{(a} \xi_{b)} = 0$.} exactly when restricted to a single geodesic. This additional specification also has the consequence that the Killing equation in a neighbourhood of a spacetime point holds up to corrections \textsl{quadratic} in the locally inertial coordinates, and holds exactly on the geodesic. 

Since the study of local horizons of \cite{Guedens:2011dy} makes use of vector fields satisfying these properties, the calculations reported in the present paper may be regarded as providing more explicit detail to the proof that such fields exist, as compared with the more sketchy treatment given in \cite{Guedens:2011dy}. NNCs are particularly well adapted to that study since the generators of the local horizon simply become coordinate curves, whilst further specialising the coordinates to be locally inertial naturally allows for the construction of an approximate boost generator. It is hoped however that locally inertial NNCs and the associated approximate Killing vectors may be of some use in other contexts as well.


\section{Construction of locally inertial null normal coordinates}
\label{LINNCs}

At a point $p$ of an $n$-dimensional spacetime, a basis $\left\{l^a, k^a, e_A^a\right\}, A=2,\ldots, n-1 $, of the tangent space is introduced, where $\left\{e_A^a\right\}$ is an orthonormal set of spatial vectors, and $l^a$ and $k^a$ are null vectors orthogonal to $e_A^a$. Furthermore, the null vectors are scaled so that their inner product equals minus one\footnote{We use the sign conventions of \cite{Misner:1974qy}, i.e. the metric signature is  $-++ \ldots$ and the Riemann tensor is defined by 
$2 \nabla_{[a} \nabla_{b]}\o_c = R_{abc}^{\ph{abc}d} \o_d$. Lower case Latin indices are used for abstract index notation, whilst Greek indices denote coordinate components. Finally, upper case Latin indices denote coordinate components on a codimension two spatial surface.}. In summary,
\beq \label{homebase}
k_a k^a=l_a l^a =k_a e_A^a=l_a e_A^a=0,\quad l_a k^a=-1 
\eeq
and 
\beq
\left(e_A\right)_a e_B^a= \d_{AB}.
\eeq

We assign the Riemann normal coordinates (RNCs, see \cite{Misner:1974qy}) $\left\{y^\a\right\} \equiv \left\{u,v,y^A\right\}$ to a point in the neighbourhood of $p$ if it lies at unit affine parameter on the geodesic through $p$ with tangent vector at $p$ given by    
$u\,l^a + v\,k^a + y^A e_A^a$. 
The components of the inverse metric are expanded as 
\beq \label{RNCg}
g^{\a\b}=\bar{\eta}^{\a\b} + \frac{1}{3} R^{\a\ph{\m}\b}_{\ph{\a}\m\ph{\b}\n}\,\,y^\m y^\n + \oythree.
\eeq   
Here, $\oythree$ denotes terms proportional to at least third powers of RNCs and $\bar{\eta}^{\a \b}$ is the flat metric in double null coordinates, namely
\beq
\bar{\eta}^{\a\b}=-2 \d^{\a}_{(u} \d^{\b}_{v)} + \sum_A \d^{\a}_{A} \d^{\b}_{A}.
\eeq   
Furthermore, throughout the text Riemann components are understood to be evaluated at the origin only\footnote{with the exception of equation (\ref{LHScomp}).}.

Let $\S$ be the $(n-2)$-dimensional spatial surface defined by the equations $u=v=0$, i.e. $\S$ is generated by geodesics with tangent at $p$ of the form $y^A e_A^a$. We now look for null vector fields spanning the planes orthogonal to $\S$; they will serve as coordinate vectors of the NNC system. Note that the parallel transport of e.g. $k^a$ along a generator of $\S$ does not in general remain orthogonal to $\S$ in a curved spacetime, while the orthogonal projection of the parallel transport of $k^a$ is not guaranteed to be a null vector. On the other hand, the gradients of $u$ and $v$ are orthogonal everywhere.      
However, they are null only at $p$, with $-\nabla^a u(p)=k^a$ and $-\nabla^a v(p)=l^a$. Therefore, we will look for linear combinations of the gradients of $u$ and $v$ that are null everywhere on $\S$. Solving a quadratic equation and choosing the appropriate sign shows that the vectors 
\beq
K^a = -\nabla^a u + K \nabla^a v, \qquad L^a = -\nabla^a v + L \nabla^a u 
\eeq   
are null everywhere on $\S$, provided that $K$ and $L$ take the values   
\beq
K=\left(g^{vv}\right)^{-1} \left[g^{uv}+\sqrt{\left(g^{uv}\right)^2-g^{uu}g^{vv}}\right] 
= -\frac{1}{2} g^{uu} + \oyfour
\eeq   
and
\beq
L=\left(g^{uu}\right)^{-1} \left[g^{uv}+\sqrt{\left(g^{uv}\right)^2-g^{uu}g^{vv}}\right] 
= -\frac{1}{2} g^{vv} + \oyfour.
\eeq   
Rescaling\footnote{Explicitly, the rescaling factor is given by 
$ \left(-L_b K^b\right)^{-1/2} = 2 \left[\left(g^{uv}\right)^2-g^{uu}g^{vv}\right]\left[g^{uv}+\sqrt{\left(g^{uv}\right)^2-g^{uu}g^{vv}}\right]/\left(g^{uu} g^{vv}\right)$.} 
the vectors so as to make their inner product equal to minus one, we obtain that for all $q \in \S$ the vectors 
\beq \label{kl}
k^a(q) = \left(-L_b K^b\right)^{-1/2} K^a, \qquad 
l^a(q) = \left(-L_b K^b\right)^{-1/2} L^a 
\eeq   
are null, orthogonal to $\S$, satisfy $l_a(q)k^a(q)=-1$ and at $p$ reduce to the coordinate vectors of $u$ and $v$. 
Anticipating the introduction of the NNC system we denote the spatial RNCs on $\S$ by
\beq \label{spaco}
x^A \equiv y^A(q).
\eeq   
The RNC components of the null vectors (\ref{kl}) are expanded in terms of the spatial coordinates (\ref{spaco}) as
\beq \label{null 1}
k^v = 1 - \frac{1}{6}R_{uAvB}\,\,x^Ax^B+\oxthree, \quad 
l^u = 1- \frac{1}{6}R_{uAvB}\,\,x^Ax^B+\oxthree,   
\eeq   
\beq
k^u = - \frac{1}{6}R_{vAvB}\,\,x^Ax^B+\oxthree, \quad 
l^v = - \frac{1}{6}R_{uAuB}\,\,x^Ax^B+\oxthree,   
\eeq   
\beq \label{null 3}
k^C =  \frac{1}{3}R_{CAvB}\,\,x^Ax^B+\oxthree, \quad 
l^C =  \frac{1}{3}R_{CAuB}\,\,x^Ax^B+\oxthree.   
\eeq   

To introduce NNCs
$\left\{ x^\a \right\} \equiv \left\{ U,V,x^A \right\} $, 
we note that a point $r$ in the neighbourhood of $p$ lies on a unique geodesic intersecting $\S$ orthogonally, at $q$ say. Let the tangent to that geodesic at $q$ be given by $U l^a + V k^a$ when $r$ lies at unit affine parameter $\z$, see figure 1. 
The NNCs of $r$ are then defined by
\beq
x^\a(r)= U\d_U^\a + V\d_V^\a + y^A(q)\d^\a_A ,
\eeq           
or in other words
\beq \label{NNC}
x^0(r)= U ,\, x^1(r)= V,\, x^A(r)=y^A(q) = x^A.
\eeq           
These comprise a locally inertial coordinate system at $p$, as we will now show (Strictly speaking, it is the coordinate system $\{(U+V)/\sqrt{2},(U-V)/\sqrt{2}, x^A\}$ which is locally inertial in the usual sense.). 
\begin{figure}[ht]
\centering
\includegraphics[width=6cm]{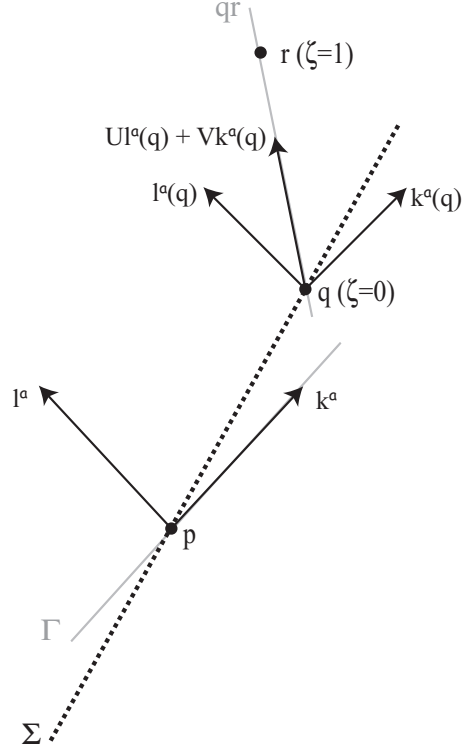}
\caption{\small The geodesic surface $\S$ in 3-dimensional spacetime (imagined to be a line segment perpendicular to the page). The null vectors $k^a$ and $l^a$ are orthogonal to $\S$ everywhere. The point $p \in \S$ is at the origin of the coordinate system and $\G$ is the null geodesic through $p$ with tangent $k^a$. At $q \in \S$ the orthogonal geodesic $qr$ has a tangent $U l^a(q)+ V k^a(q)$ such that $r$ lies at unit affine parameter $\z$. The null normal coordinates (NNCs) of $r$ are given by 
$\left\{x^\a(r)\right\}=\left\{U,V, y^A(q)\right\}$, where $y^A(q)$ are Riemann normal coordinates of $q$ on $\S$.}
\label{the surf}
\end{figure}

To find the lowest order terms of the metric components in the NNCs (\ref{NNC}), we establish the coordinate transformation between RNCs and NNCs by solving the geodesic equation perturbatively. Namely, the RNCs $y^\a(r)$ of the point $r$, lying at affine parameter $\z=1$ on the geodesic intersecting $\S$ orthogonally at $q$, are the solutions $y^\a(q, 1)$ to the equation
\beq
\frac{\mathrm{d}^2 y^\a}{\mathrm{d}\z^2}(q,\z) 
+ \G^\a_{\b \g} (q,\z) \frac{\mathrm{d} y^\b}{\mathrm{d}\z}(q,\z) \frac{\mathrm{d} y^\g}{\mathrm{d}\z}(q,\z) =0,
\eeq           
with initial conditions
\beq
u\left(q,0\right)=v\left(q,0\right)=0,\qquad
y^A\left(q,0\right) = y^A(q)= x^A 
\eeq           
and
\beq
\frac{\mathrm{d} y^\a}{\mathrm{d}\z}(q,0)= U l^\a(q) + V k^\a(q).
\eeq           
Using the expressions $\left( \ref{null 1}\right) - \left(\ref{null 3}\right)$ for $k^\a(q), l^\a(q)$ and expanding the Christoffel symbols of the RNCs in terms of the Riemann tensor \cite{Misner:1974qy}, we solve this equation to obtain the RNCs at $r$ in terms of the NNCs at $r$ and the RNC components of the Riemann tensor at $p$. However, the result shows that to first order both coordinate systems coincide, so that  $\partial y^\a / \partial x^{\b} (p) =  \d^\a_{\b}$. The tensor transformation law then tells us that at $p$ the NNC components of the Riemann tensor coincide with those in RNCs. Thus, with 
\beq
x^i \equiv U,V
\eeq
the coordinate transformation may be written as 
\beq
v(r)= V + \frac{1}{6}R_{UABi}\,\,x^A x^B x^i + \frac{1}{3}R_{UVBi}\,\,V x^B x^i +\oxfour ,
\eeq           
\beq
u(r)= U + \frac{1}{6}R_{VABi}\,\,x^A x^B x^i + \frac{1}{3}R_{VUBi}\,\,U x^B x^i +\oxfour , 
\eeq  
\beq
y^C(r)= x^C - \frac{1}{3}R_{CABi}\,\,x^A x^B x^i - \frac{1}{3}R_{CjBi}\,\, x^j x^B x^i +\oxfour.
\eeq           
Applying the tensor transformation law to (\ref{RNCg}) then obtains the NNC components of the metric, to wit 
\beq \label{metrici}
g_{UV}= -1 + \frac{1}{3}R_{UVUV}\,\,VU + \oxthree,
\eeq           
\beq 
g_{UU}= \frac{1}{3}R_{UVVU}\,\,V^2 + \oxthree,
\eeq           
\beq
g_{VV}= \frac{1}{3}R_{VUUV}\,\,U^2 + \oxthree,
\eeq           
\beq 
g_{AB}= \d_{AB} + R_{AijB} \,\, x^ix^j +
\frac{4}{3}R_{A(iC)B}\,\,x^i x^C + \frac{1}{3}R_{ACDB}\,\,x^C x^D + \oxthree
\eeq
\beq
g_{UA}= \frac{2}{3}R_{UViA}\,\,V x^i + \frac{1}{2}R_{UVBA}\,\,V x^B + \oxthree,
\eeq
\beq \label{metricf}
g_{VA}= \frac{2}{3}R_{VUiA}\,\,U x^i + \frac{1}{2}R_{VUBA}\,\,U x^B + \oxthree.
\eeq           

\subsection{Maximally simple NNCs}

Manifestly, the coordinate system defined by (\ref{NNC}) is locally inertial and null normal as defined in the introduction, but not uniquely so. An infinite number of coordinate systems with these properties exists, even if the coordinate basis $\left\{l^a, k^a, e_A^a\right\}$ is kept fixed at $p$. Since no natural scale is associated to null geodesics, we may rescale the null vectors of Eqn. (\ref{kl}) through
\beq
\tilde{k}^a(q)=f(q)\,\, k^a(q), \qquad \tilde{l}^a(q)=\frac{1}{f(q)} \,\,l^a(q)
\eeq           
where the rescaling factor $f(q)$ is a smooth function on $\S$, chosen such that $f(p)=1$. The rescaled vectors are still null, orthogonal to $\S$ and satisfy $\tilde{l}_a \tilde{k}^a=-1$. They may again be used to construct NNCs, by the same procedure as outlined above. However, if $\partial_A f (p) \neq 0$, these coordinates will no longer be locally inertial. Therefore, for a given basis $\left\{l^a, k^a, e_A^a\right\}$ at $p$, the class of locally inertial NNCs obtains from the one defined above by restricting to rescaling factors of the form 
\beq \label{rescale}
f(q) = 1 + f_{AB}\,\,x^A x^B + \oxthree, \quad f_{AB} \in \mathbb{R}.
\eeq           
Without loss of generality, the coefficients $f_{AB}$ may be taken to be symmetric in $AB$. 

Such a rescaling has no effect on the lowest order terms of the metric components except in $g_{iA}$, for which one finds
\beq 
\tilde{g}_{UA}= g_{UA} - 2f_{AB}\,\,Vx^B + \oxthree,
\qquad 
\tilde{g}_{VA}= g_{VA} + 2f_{AB}\,\,Ux^B + \oxthree.
\eeq
One may wish to simplify the form of $\tilde{g}_{UA}$, by choosing a rescaling factor with appropriate coefficients $f_{AB}$.  
However, the coefficient of the term proportional to $Vx^B$ in $g_{UA}$ is given by $ \frac{1}{2} \,\, R_{UVBA}$. This is \textsl{anti}symmetric in $AB$, so the term proportional to $Vx^B$ cannot be removed by a rescaling of the form (\ref{rescale}). Similar remarks of course hold for $g_{VA}$. Accordingly, the locally inertial NNC system defined by (\ref{NNC}) is the one in which the expression for the metric components is as simple as possible.


\section{Construction of a local, approximate Killing vector} 
\label{LKV}

In a neighbourhood of a spacetime point $p$, we define a vector field $\xi^a$ by taking its components in the NNCs (\ref{NNC}) to be a power series
\beq \label{xicon}
\xi^\a = V\d_V^\a -U\d_U^\a + \oxtwo,
\eeq
such that the $\oxtwo$ terms vanish in Minkowski space. In Minkowski space, this defines the Killing vector that generates boosts in the $UV$ plane, with fixed point at $p$. In curved spacetime no Killing vectors are guaranteed to exist, but $p$ is still a fixed point of $\xi^a$ at which the Killing equation is satisfied, and furthermore $\nabla_{(\a} \xi_{\b)}=\oxone$. 

We will impose a restriction on $\xi^a$ that determines many of the higher order terms in its expansion.  
For a true Killing vector, combining different permutations of the covariant derivative of the Killing equation $\nabla_{(a} \xi_{b)} = 0$ results in the Killing identity
\beq \label{KI}
\nabla_r \nabla_s \xi_a - R^m{}_{rsa} \xi_m = 0. 
\eeq
For approximate Killing vectors of the form (\ref{xicon}), the LHS of (\ref{KI}) is at best linear in the coordinates when evaluated in the neighbourhood of a point in a general spacetime. However, $\xi^a$ may be chosen so as to make the LHS of (\ref{KI}) vanish when restricted to a single curve, as we will now show.
  
We begin by writing out the LHS of (\ref{KI}) in terms of partial derivatives and Christoffel symbols, as
\beq 
\label{LHScomp}
\nabla_\r \nabla_\s \xi_\a - R^\m{}_{\r\s\a} \xi_\m =  & \p_\r \p_\s \xi_\a 
+ \left( -\p_\r \G^\m_{\ph{\m}\s\a} + 2 \G^\n_{\ph{\m}\r(\s}\G^\m_{\ph{\m}\a)\n} \right)\xi_\m 
- \left(\G^\m_{\ph{\m}\r\s} \p_\m\xi_\a + 2 \G^\m_{\ph{\m}\a(\r} \p_{\s)}\xi_\m \right) 
- R^\m{}_{\r\s\a} \xi_\m .
\eeq
This is an equation valid in any coordinate system and it is the only equation in this paper where Riemann components and Christoffel symbols take values in a spacetime neighbourhood, not just at $p$.   Next we write the covariant components of $\xi^a$ in NNCs as the power series
\beq
\label{local_killing_NNC}
\xi_\a = U\d_\a^{V} - V \d_\a^{U}
+ \frac{1}{2} C_{\r\s\a} x^\r x^\s
+ \frac{1}{3!} D_{\r\s\m\a} x^\r x^\s x^\m + \frac{1}{4!} E_{\r\s\m\n\a} x^\r x^\s x^\m x^\n + \oxfive,
\eeq
where the coefficients $C_{\r\s\a}$, $D_{\r\s\m\a}$ etc. are symmetric in all but their last index.   
Lastly, let $\G$ be the null geodesic through $p$ with tangent $k^a$. We will evaluate the NNC components of the Killing identity term (\ref{LHScomp}) on $\G$, using the approximate Killing vector (\ref{local_killing_NNC}). Since $\G$ is simply a coordinate curve, with points on $\G$ having NNCs $(0,V,0,0,...)$, the NNC components of (\ref{LHScomp}) on $\G$ reduce to a power series in $V$. Making use of the fact that at $p$ the coordinate system is locally inertial, we find
\beq
\label{powers of V}
\left. \left( \nabla_\r \nabla_\s \xi_\a - R^\m{}_{\r\s\a} \xi_\m\right)\right \vert_{\G} =
  C_{\r\s\a} 
  + \Bigl( D_{V \r \s \a} + \la \r\s\a \ra - \la  \a\r\s \ra - \la \a\s\r \ra
 - \partial_\r \Gamma_{V\s\a} - R_{V\r\s\a} \Bigr)\, V 
  + \oVtwo. 
\eeq
As in the rest of the text (with the exception of (\ref{LHScomp})) the Riemann components and Christoffel symbols on the RHS of (\ref{powers of V}) are evaluated at $p$. Furthermore, we defined $\G_{\a\b\g} \equiv g_{\a\d} \G^\d_{\ph{\d}\b\g}$ and for brevity we introduced the notation 
\beq
\la \r\s\a \ra \; \equiv \; 2 \partial_V \Gamma^{[V}_{\ph{[V}\r\s}\d^{U]}_\a. 
\eeq

Now we are ready to impose that the Killing identity hold on $\G$, which amounts to putting the RHS of (\ref{powers of V}) to zero, order by order. 
At zeroth order this requires
\beq 
C_{\r\s\a} = 0,
\eeq
and as a consequence the quadratic term in $\xi^a$ vanishes, i.e. $\xi^\a = V\d_V^\a -U\d_U^\a + \oxthree$. A further consequence is that the Killing equation is satisfied in a neighbourhood of $p$ up to terms quadratic in the coordinates, $\nabla_{(\a} \xi_{\b)}=\oxtwo$. Given the linear terms in $\xi^a$, this is as good as it gets: no choice of third order terms can reduce the order of the Killing equation further. 

At first order, imposing the Killing identity on $\G$ requires
\beq \label{third order coeffs}
D_{V \r \s \a} = - \la \r\s\a \ra + \la  \a\r\s \ra 
+ \la \a\s\r \ra
+ \partial_\r \G_{V\s\a} + R_{V\r\s\a}. 
\eeq
Note that this does not fully determine the third order terms in $\xi^\a$, but only those that are at least linear in $V$. To write the $D$-coefficients more explicitly, we note that at the origin of any locally inertial coordinate system (and for the NNCs (\ref{NNC}) in particular) it holds that 
\beq \label{licresult}
\p_\a\G_{\b\g\d} = -\frac{1}{2} \p_\a\p_\b \, g_{\g\d} +  \p_\a\p_{(\g} \, g_{\d)\b}.
\eeq
In addition, we may express Riemann components at the origin of a locally inertial coordinate system in terms of second derivatives of the metric. 
Substituting these expressions into (\ref{third order coeffs}) results in
\beq \label{result}
 D_{V \r \s \a} =  - \la \r\s\a \ra + \la \a\r\s \ra + \la \a\s\r \ra
+ \p_\r\p_\s \, g_{\a V} + \frac{1}{2} \p_\a\p_V \, g_{\r\s}  -  \p_V\p_{(\r} \, g_{\s)\a}, 
\eeq
with the brackets given by 
\beq \label{brac}
\la \r\s\a \ra \; = \;  
 \left( - \frac{1}{2} \p_V\p_V \, g_{\r\s} +  \p_V\p_{(\r} \, g_{\s)V}\right) \d^V_\a 
+ \left(  \frac{1}{2} \p_V\p_U \, g_{\r\s} -  \p_V\p_{(\r} \, g_{\s)U}\right) \d^U_\a .
\eeq
Clearly, the bracket $\la \r\s\a \ra$ is symmetric in the index pair $\r \s$. It is then manifest that the RHS of (\ref{result}) is also symmetric in $\r \s$, as is necessary for consistency with the symmetry of $D_{V \r \s \a}$.  
We remind the reader that the second derivatives of the metric are to be evaluated at $p$. These derivatives can be read off directly from the metric components listed in (\ref{metrici})-(\ref{metricf}) and they are always proportional to a component of the Riemann tensor. Accordingly, all the coefficients  $D_{V \r \s \a}$ are required to be a certain (sum of) Riemann component(s) evaluated at $p$.   
 
Similarly, at second order the Killing identity on $\G$ requires the coefficients $E_{VV \r \s \a}$ to be given in terms of partial derivatives of the Riemann components. At the next order, coefficients $F_{VVV \r \s \a}$ are required to be a sum of second derivatives of Riemann components and products of Riemann components, and so on. Hence we see that the Killing identity may be satisfied exactly on the null geodesic $\G$ through $p$. This also has the consequence that $\xi^a$ satisfies the Killing equation exactly on $\G$. Furthermore, as shown in \cite{Guedens:2011dy}, $\xi^a$ is null on $\G$ and has the same proportionality to an affine tangent as has a Killing vector generating a Killing horizon, namely $\xi^a \vert_{\G} = V k^a$. 

\subsection{Placing the fixed point to the past of the origin}
 
In the treatment of \cite{Guedens:2011dy}, the origin of the coordinate system was placed on a local causal horizon's terminal surface, such as the surface $\S$ we have considered here. However, it turns out to be more natural thermodynamically if the fixed point of the approximate Killing vector is placed slightly to the past of the terminal surface (see \cite{Guedens:2011dy} for further details). We would therefore like to specify an approximate Killing vector that still satisfies the Killing identity on $\G$, but with the fixed point now chosen to be a point on $\G$ with NNCs $(0,V_0, 0,0,...)$, slightly to the past of $p$.
Instead of (\ref{local_killing_NNC}) we expand $\xi_\a$ in powers of $\tx^\a \equiv x^\a-V_0\d^\a_{V}$, as 
\beq
\label{shifted_NNC}
\xi_\a = U\d_\a^{V} - \tV \d_\a^{U}
+ \frac{1}{2} C_{\r\s\a}\tx^\r\tx^\s 
+ \frac{1}{3!} D_{\r\s\m\a} \tx^\r\tx^\s\tx^\m + \oxfour.
\eeq
For simplicity we will merely demand the Killing identity on $\G$ to hold at the linear order. Then a straightforward calculation reveals that the $D$-coefficients remain unchanged, but the $C$-coefficients change to 
\beq 
C_{\r\s\a} =  \Bigl(- \la \r\s\a \ra + \la  \a\r\s \ra 
+ \la \a\r\s \ra \Bigr)\,V_0. 
\eeq
This does not mean however that the approximate Killing vector picks up a quadratic term, since $V_0$ is regarded as a first order quantity.   


\acknowledgments
The author would like to thank Ted Jacobson and Sudipta Sarkar for valuable discussions.

\appendix
\section{A generalisation of Fermi-Walker transport}

As is well known, the inner product of a pair of vectors is preserved under parallel transport along a curve. If the curve is not null, any vector field defined along the curve may be uniquely decomposed into a part tangential to the curve and a part orthogonal to it. Since an affine tangent to a geodesic is parallel transported into itself, this implies that along a non-null geodesic the tangential and orthogonal parts of a parallel transported vector are parallel transported separately.  

Fermi-Walker (FW) transport \cite{Misner:1974qy} generalises the notion of parallel transport such that the above features survive along non-geodesic curves. That is, FW transport reduces to parallel transport along a geodesic, the inner product of a pair of FW transported vectors is preserved and the tangential and orthogonal parts of a vector that is FW transported along a non-null curve are FW transported separately. 

The present aim is to define a notion of transport across a multidimensional surface that again shares these features with FW transport. This will be achieved by generating the surface by a family of curves that emanate from a point and by specifying how a vector is transported along each curve. Before proceeding, it will be instructive to give a definition of FW transport along non-null curves\footnote{On null geodesics, FW transport reduces to parallel transport. Non-geodesic null curves will not be considered.}, as this will allow for a natural extension to the multidimensional case. 

Let $T$ be the projector\footnote{In this appendix, the abstract indices denoting a tensor will often be omitted when there is no cause for confusion.} onto a non-null curve and $N=1-T$ the projector onto its orthogonal complement. A vector $w$ defined along the curve may be decomposed as   
\beq
w= Tw +Nw.
\eeq
If $w$ and $x$ are vectors which are FW transported along the curve, their tangential and orthogonal parts should be FW transported separately. Therefore, the inner products of their tangential parts and of their orthogonal parts should both be preserved separately.  
For the tangential parts this reads 
\beq \label{FWtang}
0 & = & \left[ \left(Tw\right)_a \left(Tx\right)^a \right]^{\boldsymbol{.}} \nonumber \\
  & = & \left[\left(Tw\right)^{\boldsymbol{.}}\right]_a \left(Tx\right)^a  
 + \left(Tw\right)_a  \left[\left(Tx\right)^{\boldsymbol{.}}\right]^a,
\eeq
where covariant differentiation along the curve was denoted by a dot.
Allowing for arbitrary vectors at an initial point on the curve, this condition can only be satisfied provided that the covariant derivatives of the tangential parts are orthogonal vectors. Therefore, we impose
\beq \label{cond1}
\left(Tw\right)^{\boldsymbol{.}} = N \left(Tw\right)^{\boldsymbol{.}} ,
\eeq
and similarly for $x$. The latter condition is satisfied if and only if the norm of $Tw$ remains constant. Another statement   
equivalent to (\ref{cond1}) is that $Tw$ is parallel transported along the curve, according to the derivative operator compatible with the one dimensional metric induced on the curve.
Similar to the condition imposed on a tangential part $Tw$, we ensure that the inner product of orthogonal parts $Nw$ and $Nx$ is preserved under FW transport, by imposing that the covariant derivative of an orthogonal part $Nw$ is tangential. This may be written as
\beq \label{cond2}
\left(Nw\right)^{\boldsymbol{.}} = T \left(Nw\right)^{\boldsymbol{.}} .
\eeq

The condition (\ref{cond1}) together with the $(n-1)$ conditions (\ref{cond2}) can be summarized by a single differential vector equation. For this purpose, we parametrize the curve such that the tangent $e$ has a constant norm, given by $e_a\,e^a=E$. A standard choice would be to parametrize by proper length or time, for which we have $E=\pm 1$. In abstract index notation, the covariant derivative along the curve is defined by $``."=e^a \nabla_a$ and the tangential projector is given by $T^a_{\ph{a}b}= E^{-1}e^a e_b$. 
Thus the tangential part of $w$ is written as $Tw= \left(E^{-1} e^b w_b\right)\,e$. Using these expressions and substituting (\ref{cond1}) into (\ref{cond2}), we find that a vector can be defined to be FW transported if it is a solution to the equation  
\beq \label{FW}
\dot{w} = E^{-1} \left[ \left( e^b w_b\right) \, \dot{e} - \left(\dot{e}^b w_b \right) \, e \right]. 
\eeq

We will now define a vector field on an $m$-dimensional (non-null) surface $\S$, starting from a vector at a given point on the surface, and in such a way that the definition reduces to FW transport for the case $m=1$. An $m$-dimensional projector $T$ and an $(n-m)$-dimensional projector $N$ are introduced, projecting onto the surface and its orthogonal complement respectively. The metric $T_{ab}$ induced on the surface is obtained from the projection operator $T^a_b$ by lowering one index by the full spacetime metric. 
Furthermore, the surface is regarded as generated by a family of curves that emanate from a given base-point $p$. On a sufficiently small surface, every point (other than $p$) will lie on a unique curve in this family. We then focus on one particular curve, again denoting covariant differentiation along the curve by a dot.

Starting with a vector $w(p)$ at $p$, the transport of $w(p)$ along the curve is defined by the $m$ conditions
\beq \label{mcond1}
\left(Tw\right)^{\boldsymbol{.}} = N \left(Tw\right)^{\boldsymbol{.}} ,
\eeq
together with the $(n-m)$ conditions
\beq \label{mcond2}
\left(Nw\right)^{\boldsymbol{.}} = T \left(Nw\right)^{\boldsymbol{.}} ,
\eeq
in close analogy to the conditions (\ref{cond1}) and (\ref{cond2}) for FW transport. That is to say, $w$ is defined along the curve by requiring the covariant derivative of its tangential part to be orthogonal and the covariant derivative of its orthogonal part to be tangential. Condition (\ref{cond1}) may again be interpreted as stating that the tangential part of $w$ is parallel transported along the curve, only now according to the derivative operator compatible with the induced metric $T_{ab}$ on the surface. By construction, inner products of transported vectors are preserved and tangential and orthogonal parts of a transported vector are transported separately. In particular, if $w(p)$ is orthogonal to $\S$ then the transport of $w(p)$ will be orthogonal to $\S$ everywhere and will have a fixed norm\footnote{In this case, it is interesting to note that the transported vector may be regarded as obtained from a continuous application of (full metric) parallel transport followed by orthogonal projection. This is easily seen by writing out components of the transported vector an infinitesimal distance away from $p$ and comparing with the components of the orthogonal projection of the parallel transport.}. 

Repeating this for all curves through $p$ in the family that was chosen to generate $\S$ then obtains a vector field $w(q)$ on $\S$. Clearly, the definition of the transported vector field depends on the family of curves that generate $\S$. For definiteness, we choose the family of surface geodesics through $p$ associated to the induced metric $T_{ab}$. Since a different choice of base-point on $\S$ leads to a different family of surface geodesics, our definition of surface transport is subject to the choice of base-point $p$.

As in the case of FW transport, the defining conditions (\ref{mcond1}) and (\ref{mcond2}) may be combined into a single equation. To this end, we introduce a tangential basis $\{e_A(p)\}, A= 1,\dots,m$ at $p$ and construct tangential basis vectors $e_A(q)$ everywhere on $\S$ as the solutions to (\ref{mcond1}) with initial condition $e_A(p)$. In other words, $e_A(q)$ is the parallel transport of $e_A(p)$ along the surface geodesics, according to the derivative operator compatible with the induced metric on the surface. As such, $\dot{e}_A$ is orthogonal to $\S$ and the inner products $g(e_A,e_B) = T_{AB}$ are constant along $\S$. An orthonormal frame at $p$ would provide a natural initial condition to give rise to such a tangential basis. Next, writing  
$ Tw  = w^A \,e_A$, we note that the tangential components can be expressed as $w^A=T^{AB} e_B^b w_b$, where $T^{AB}$ is the inverse of the matrix $T_{AB}$. Upon substitution of (\ref{mcond1}) into (\ref{mcond2}), a straightforward manipulation then obtains 
\beq \label{transport}
\dot{w} = T^{AB} \left[ \left(e_B^b w_b\right) \dot{e}_A - \left(\dot{e}_B^b w_b \right) \,e_A \right]. 
\eeq
For the case $m=1$, this equation manifestly reduces to the defining equation (\ref{FW}) of FW transport.  
Furthermore, (\ref{transport}) implies the $n$ conditions (\ref{mcond1}) and (\ref{mcond2}) and uniquely determines a vector $w(q)$ along $\S$, once $w(p)$ is specified. In particular, it may be checked that the vector fields $k^a(q)$ and $l^a(q)$ defined in (\ref{kl}) are the unique solutions to (\ref{transport}) for the given initial vectors $k^a(p)$ and $l^a(p)$.

\end{document}